\documentclass[letterpaper,12pt,preprint]{aastex}

\def\solmass {\,$\hbox{M}_\odot$}

\def\NTDP {N$_2$D$^+$}
\def\TCO {$^{13}$CO}
\def\CEO {C$^{18}$O}
\def\CSO {C$^{17}$O}
\def\kms {km\,s$^{-1}$}

\shorttitle{Substructre in Starless Cores II}
\shortauthors{Schnee et al.}

\begin{document}

\title{An Observed Lack of Substructure in Starless Cores II:
  Super-Jeans Cores}

\author{Scott Schnee\altaffilmark{1}, Sarah Sadavoy\altaffilmark{2,3},
  James Di Francesco\altaffilmark{2,3}, Doug
  Johnstone\altaffilmark{2,3}, Lisa Wei\altaffilmark{4,5}}

\email{sschnee@nrao.edu}

\altaffiltext{1}{National Radio Astronomy Observatory, 520 Edgemont
  Road, Charlottesville, VA 22903, USA}
\altaffiltext{2}{National Research Council Canada, Herzberg Institute
  of Astrophysics, 5071 West Saanich Road Victoria, BC V9E 2E7,
  Canada}
\altaffiltext{3}{Department of Physics \& Astronomy, University of
  Victoria, Victoria, BC, V8P 1A1, Canada}
\altaffiltext{4}{Harvard-Smithsonian Center for Astrophysics, 60
  Garden Street, Cambridge, MA 02138, USA}
\altaffiltext{5}{Atmospheric and Environmental Research, 131 Hartwell
  Avenue, Lexington, MA 02421, USA}

\begin{abstract}

We present SMA and CARMA continuum and spectral line observations of
five dense cores located in the Perseus and Ophiuchus molecular clouds
whose masses exceed their thermal Jeans masses.  Three of these cores
have previously been identified as being starless and two have been
classified as being possibly protostellar.  We find that one core is
certainly protostellar.  The other four cores, however, are starless
and undetected in both \CEO\ and 1.3\,mm continuum emission.  These
four starless cores have flat density profiles out to at least
$\sim$0.006\,pc, which is typical for starless cores in general.
Density profiles predicted by some collapse models, especially in the
early stages of infall, are consistent with our observations.
Archival data reveal that these starless cores have significant
non-thermal support against collapse, although they may still be
unstable.
\end{abstract}


\section{Introduction}

Recent interferometric observations of supposedly ``starless'' cores
in nearby molecular clouds have found several hidden protostars
\citep[e.g.,][]{Pineda11, Dunham11, Schnee12, Chen12}.  These
observations, however, almost never reveal substructures or evidence
for fragmentation in starless cores \citep{Olmi05, Schnee10}, with
some notable exceptions, such as R CrA SMM 1A \citep{Chen10} and L183
\citep{Kirk09}.  The lack of observed substructure in starless cores
may imply that stellar multiplicity begins during the protostellar
stage.  For instance, in the disk fragmentation theory of multiple
star formation, a massive accretion disk around a protostar can become
unstable and fragment, creating a binary or higher-order system
\citep{Adams89, Bonnell94}.  Alternatively, turbulence in starless
cores may be forming the seeds of multiplicity \citep{Fisher04,
  Goodwin04, Goodwin07}, but these seeds may be below the threshold of
detectability, as recently suggested by \citet{Offner12}.

In this paper, we present interferometric observations of five cores
whose masses derived from dust emission at 850\,\micron\ exceed their
respective thermal Jeans masses by at least a factor of 2.  As such,
they are likely to be in the process of forming protostars and/or
fragmenting.  These cores, given the label ``super-Jeans'', were
identified by \citet{Sadavoy10b} and thus make promising targets in
the search for substructure in starless cores.

\section{Observations} \label{OBSERVATIONS}

\subsection{Previous Observations and Sample Definition} \label{SAMPLE}

The cores in this study were previously observed at 850\,\micron\ with
the Submillimeter Common-User Bolometer Array (SCUBA) on the James
Clerk Maxwell Telescope (JCMT) \citep{DiFrancesco08} and at mid- to
far-infrared wavelengths with the {\it Spitzer Space Telescope}
\citep{Evans03, Evans09}.  Core properties such as mass and size were
derived from the SCUBA maps by \citet{Sadavoy10a}, and the
classification of each core (i.e., starless, protostellar, or
``undetermined'') was made from 3.6-70\,\micron\ {\it Spitzer} maps
\citep{Sadavoy10b}.  \citet{Sadavoy10b} estimated the stability of
cores by comparing the Jeans mass for a thermally supported sphere
against the core mass derived from dust emission.  The Jeans mass is
given by
\begin{equation}
\label{JEANSMASS}
M_J = 1.9 \left( \frac{T_d}{10\,\mathrm{K}} \right) \left(
\frac{R_J}{0.07\,\mathrm{pc}} \right) M_\odot
\end{equation}
while the dust-derived mass is given by
\begin{equation}
\label{DUSTMASS}
M = 0.074 \left( \frac{S_{850}}{\mathrm{Jy}} \right) \left(
\frac{d}{\mathrm{100\,pc}} \right)^2 \left(
\frac{\kappa_{850}}{\mathrm{0.01\,cm^2\,g^{-1}}} \right)^{-1}  \times
 \left[\mathrm{exp} \left( \frac{\mathrm{17\,K}}{T_d} \right) - 1
  \right] M_\odot
\end{equation}
The dust temperatures ($T_d$) assumed by \citet{Sadavoy10b} are 15\,K
in Ophiuchus and 11\,K in Perseus, based on NH$_3$ surveys of these
clouds by \citet{Friesen09} and \citet{Rosolowsky08}, respectively.
The Jeans radius ($R_J$) and 850\,\micron\ flux ($S_{850}$) are
estimated from the SCUBA observations, the distances assumed are shown
in Table \ref{PROPERTIESTABLE}, and the dust opacity at
850\,\micron\ was assumed to be 0.01 cm$^{2}$ g$^{-1}$ by
\citet{Sadavoy10b}.

Using the data and equations above, \citet{Sadavoy10b} identified 17
candidate starless cores in nearby molecular clouds whose dust-derived
masses exceeded their respective Jeans masses by more than a factor of
2.  Based on a simple Jeans analysis, such cores would be expected
either to have significant non-thermal support or else to be
collapsing and possibly fragmenting.  Of these 17 super-Jeans cores,
we chose 5 for follow-up observations with the Submillimeter Array
(SMA).  We chose to observe all 3 of the super-Jeans cores that
retained their ``starless'' classification (Oph-2, Per-2, and Per-6)
and 2 of the super-Jeans cores reclassified as ``undetermined'' (Oph-1
and Per-8) by \citet{Sadavoy10b}.  Per-2 and Per-8 were also observed
with CARMA.  Basic information on these 5 cores is shown in Table
\ref{PROPERTIESTABLE}, and more information about these cores and
their properties were described by \citet{Sadavoy10a, Sadavoy10b}.

\subsection{SMA Calibration}  \label{SMA}

Spectral line observations were taken with the eight-element
Submillimeter Array (SMA) \citep{Ho04} in June 2010 and October 2010
in the compact-north and compact configurations,
respectively. Baselines range from 16\,m to 77\,m for both
configurations, providing sensitivity to spatial scales up to
$\sim$8\arcsec\ and a resolution of $\sim$3\arcsec\ FWHM. The sources
were grouped together by RA and observed together in single tracks
(Oph-1 and Oph-2 in June; Per-2, Per-6, and Per-8 in October) each
about 10 hours in length (see Table \ref{OBSPROPTABLE} for
details). The half-power beam width of the 6\,m antennas is
54\arcsec\ at 230\,GHz, and all sources were observed with single
pointings.

The receivers were tuned to a rest frequency of 230.538\,GHz, Doppler
tracked to a recessional velocity (V$_{\rm LSR}$ of 3.95\,\kms\ and
7.8\,\kms\ for the June and October observations, respectively). The
correlator was configured to observe the 1.3\,mm continuum with 72
windows, each 104\,MHz wide, with a total band width of 7\,GHz in the
upper and lower sidebands. We also configured four additional windows
to observe the $^{12}$CO(2--1) and \TCO(2--1) lines with 406\,kHz
(0.5\,\kms) resolution, the \CEO\ line with 203\,kHz (0.25\,\kms)
resolution, and the \NTDP\ (3-2) line with 813\,kHz (1\,\kms)
resolution (see Table \ref{OBSTABLE}).  The \TCO\ (2--1) and
\NTDP\ (3-2) lines were only marginally detected towards one core
(Per-8), so neither transition is discussed further in this paper.
\CEO\ (2--1) was detected only towards Per-8.

The data were reduced using the MIR package. We flagged the data for
bad channels, antennas, weather, and pointing. Radio pointing was done
at one hour after sunset for each track. We calibrated the bandpass
with a bright quasar, 3C~454.3, for both tracks. Atmospheric
calibration was done with observations of the quasars 1625-254 and
1517-243 in the June track and 0036+323 and 0319-415 in the October
track every 15 minutes. Absolute flux calibration was done using
observations of Uranus.

\subsection{CARMA Calibration} \label{CARMA}

Spectral line observations in the 1\,mm window were obtained with
CARMA, a 15-element interferometer consisting of nine 6.1-meter
antennas and six 10.4-meter antennas.  The CARMA observations were
taken in D-array for both observations, with baselines ranging from
11\,m to 150\,m, providing sensitvity to spatial scales up to
$\sim$12\arcsec\ and a resolution of $\sim$2\arcsec\ FWHM.  The
sources Per-2 and Per-8 were observed in single pointings in separate
tracks in May 2011 and June 2011, respectively, each for about 4.5
hours (see Table \ref{OBSPROPTABLE} for details).

The receivers were tuned to a rest frequency of 230.538\,GHz and
Doppler tracked to a recessional velocity (V$_{\rm LSR}$) of
7.8\,\kms.  The correlator was configured to observe the 1.3\,mm
continuum with 7 bands, each with an upper and lower sideband, with
495\,MHz band width with 39 channels per band, providing a total band
width of 7\,GHz.  One band was configured to observe \CSO\ (2-1) in
the lower sideband and CO (2-1) in the upper sideband, each with
781\,kHz spectral resolution ($\sim$1\,\kms) and 123\,MHz band width
(See Table \ref{OBSTABLE}).

The data were reduced using the MIRIAD package \citep{Sault95}. We
flagged the data for bad channels, antennas, weather, and pointing.
Radio pointing was done at the beginning of each track and pointing
constants were updated at least every two hours thereafter, using
either radio or optical pointing routines \citep{Corder10}.  We
calibrated the bandpass and gains using observations of the bright
quasar 3C84, which was observed for 3 minutes out of each 21 minute
source-calibrator cycle.  Absolute flux calibration was done using
observations of Mars and the consistency of flux measurements with the
SMA data set.

Due to instrumental problems, the upper sideband of the spectral line
band was lost, so our observations of CO (2-1) come from the SMA only.
\CSO\ (2--1) was detected only towards Per-8.

\subsection{Imaging} \label{Imaging}

We imaged the SMA and CARMA continuum and line emission from each
source using MIRIAD \citep{Sault95}, with natural weighting and an
additional weighting in inverse proportion to the noise as estimated
by the system temperature. Each data cube was cleaned to a cutoff of
2\,$\sigma$ in the residual image.  The 1.3\,mm continuum data for the
source Per-8 came from both CARMA and the SMA, while all other maps
have data from only one array.  The combined CARMA+SMA 1.3\,mm
continuum map of Per-8 has an rms of 6 mJy\,beam$^{-1}$ and a beam
size of 2.7\arcsec$\times$2.0\arcsec.

\section{Analysis} \label{ANALYSIS}


\subsection{Per-8} \label{DETECTIONS}

Continuum emission at 1.3\,mm is detected from Per-8, with a peak flux
of 380\,mJy beam$^{-1}$ and an integrated flux of 750\,mJy.  The peak
of the dust emission seen with SMA and CARMA is at (J2000) 3:32:17.923
+30:49:47.869, and this is also the peak of the detected \CSO\ (2-1)
and \CEO\ (2--1) integrated intensity.

\citet{Sadavoy10a} initially classified Per-8 as ``starless,'' after
rejecting a nearby compact source of mid-infrared emission detected by
{\it Spitzer} in the IRAC and MIPS bands (SSTc2dJ033218.0+304947) as
non-protostellar.  This compact source has colors similar to those of
star-forming galaxies \citep[e.g., using the prescription
  from][]{Gutermuth08} rather than the colors expected for embedded
protostars \citep[e.g., following][]{Evans09}.  Since such color
analyses are not perfect and may throw out legitimate protostellar
sources, \citet{Sadavoy10b} reclassified Per-8 as ``undetermined.''
Indeed, Per-8 was classified as a Class 0 protostar by
\citet{Hatchell09}, who identified an outflow from the core.
Near-infrared maps of the Perseus B1 region also show shocked emission
from an outflow driven by Per-8 \citep{Walawender09}.

\begin{figure}
\epsscale{1.0} 
\plotone{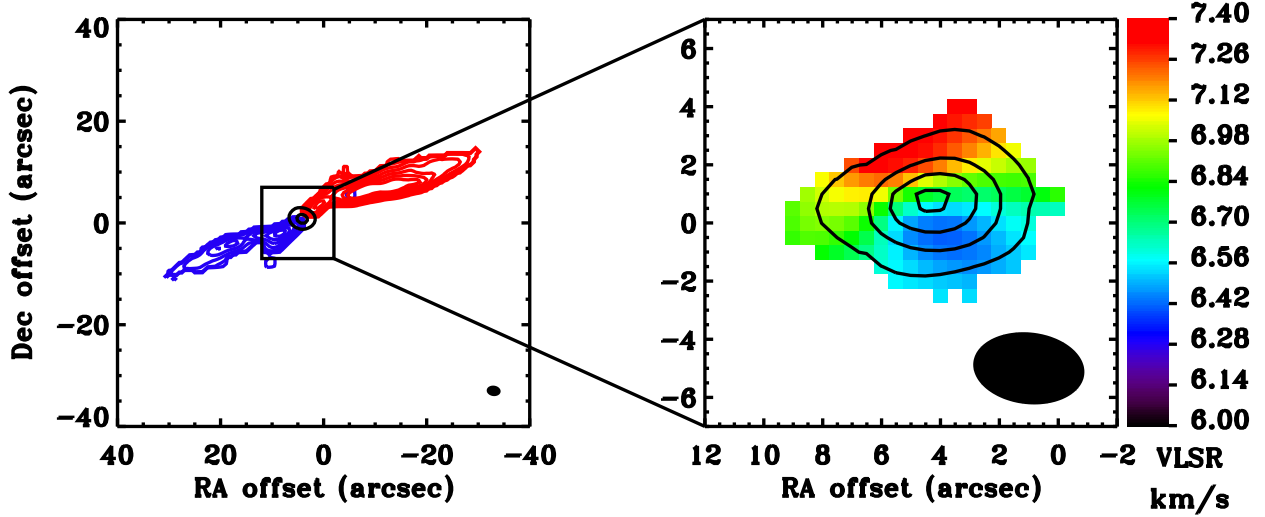}
\caption{Interferometer data of Per-8.  ({\it left}) Black contours
  show the 1.3\,mm continuum emission from Per-8, at values of
  0.1\,Jy\,beam$^{-1}$ and 0.3\,Jy\,beam$^{-1}$.  Red and blue
  contours show the redshifted and blueshifted CO (2--1) integrated
  intensity, beginning at a value of 3\,K\,\kms\ and increasing with
  steps of 3\,K\,\kms.  Redshifted emission is integrated over LSR
  velocities between 9\,\kms\ and 18\,\kms, and blueshifted emission
  is integrated over LSR velocities between -2\,\kms\ and 5\,\kms. The
  synthesized beam of the continuum emission is shown in the bottom
  right corner of the panel.  ({\it right}) Black contours show the
  the integrated intensity of the \CEO\ (2--1) emission beginning at a
  value of 2\,K\,\kms\ and increasing with steps of 2\,K\,\kms.  Color
  shows the centroid velocity from Gaussian fits to the \CEO\ (2--1)
  spectra.  The synthesized beam of the \CEO\ (2-1) data is shown in
  the bottom right corner of the panel.  Note that the synthesized
  beams of continuum (2.7\arcsec\ $\times$ 2.0\arcsec) and CO
  (3.7\arcsec\ $\times$ 2.4\arcsec) data differ because the former
  come from CARMA and the SMA, while the latter only come from the
  SMA.
\label{SMAMAPS}}
\end{figure}

\begin{figure}
\epsscale{1.0} 
\plotone{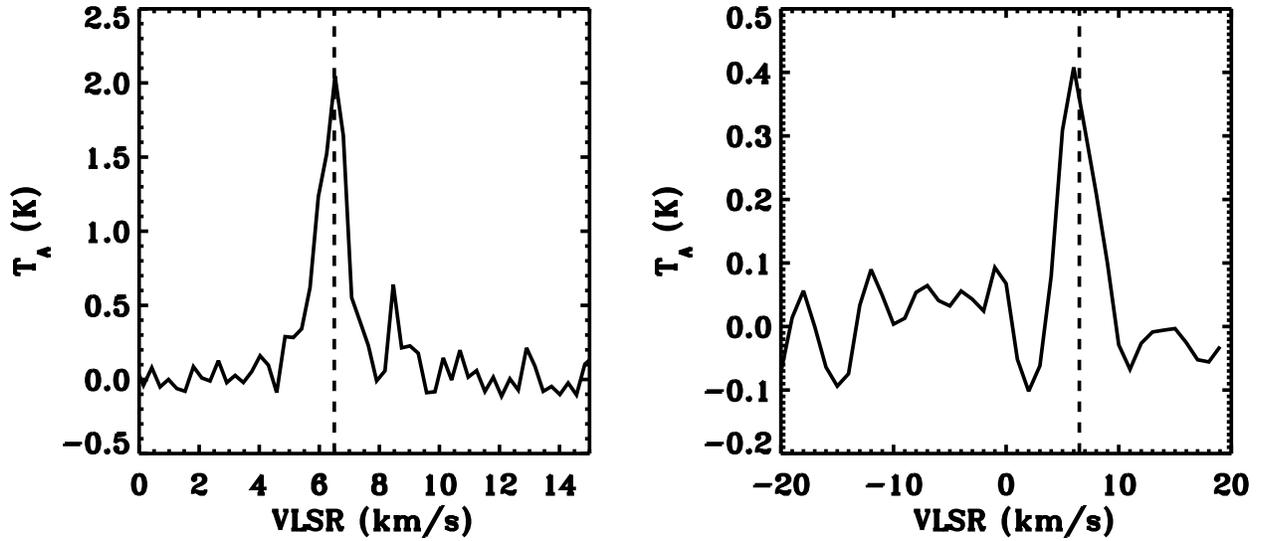}
\caption{Spectra towards Per-8.  ({\it left}) \CEO\ (2-1) spectrum
  taken at the peak of the \CEO\ (2-1) integrated intensity (J2000
  3:32:17.87, +30:49:47).  ({\it right}) \CSO\ (2-1) spectrum taken at
  the peak of the \CSO\ (2-1) integrated intensity (same location as
  for \CEO\ (2-1)).  The vertical dashed line in both panels shows a
  velocity of 6.5\,\kms\ and is plotted to guide the eye and show that
  the \CEO\ (2-1) and \CSO\ (2-1) spectra peak at the same velocity.
\label{SPECTRA}}
\end{figure}

\begin{figure}
\epsscale{1.0} 
\plotone{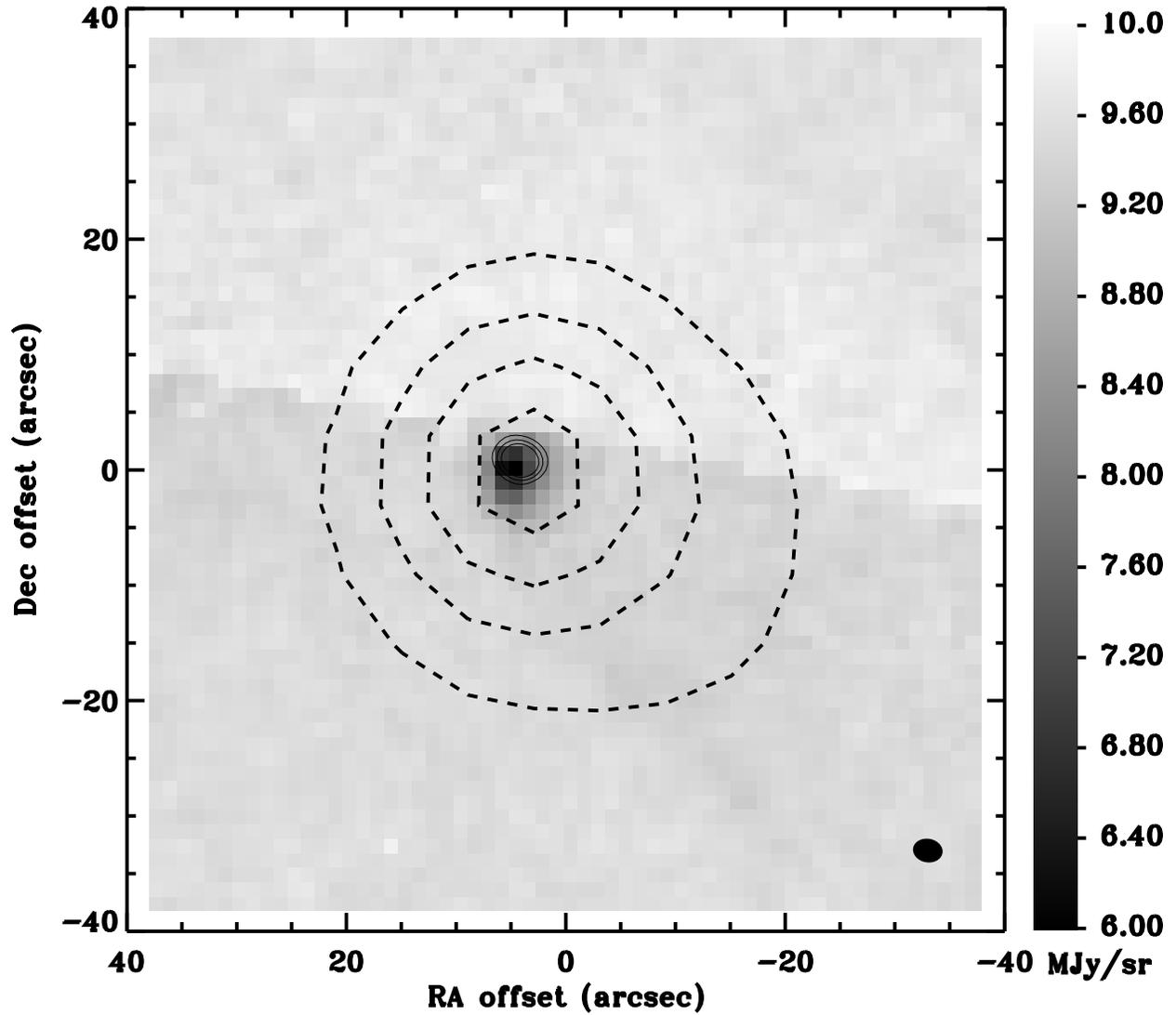}
\caption{Continuum data of Per-8.  Greyscale shows the {\it Spitzer}
  IRAC Band 4 (8\,\micron) emission from the region around Per-8.  The
  1.3\,mm continuum emission detected with SMA and CARMA (thin solid
  contours at 100\,mJy/beam, 150\,mJy/beam, and 200\,mJy/beam) are
  coincident with the 8\,\micron\ emission.  The thick dashed contours
  show the 850\,\micron\ emission (at 0.5, 1.0, 1.5, and 2.0 Jy/beam)
  from Per-8 detected with SCUBA \citep{DiFrancesco08}.
\label{IRACMAP}}
\end{figure}

Given the previous observations of this source, it is not surprising
that we detect a collimated CO(2--1) outflow from Per-8.  The outflow
is centered on the 1.3\,mm continuum source, which is also coincident
with the compact source detected with {\it Spitzer}.  In Figure
\ref{SMAMAPS}, we show the outflow, the 1.3\,mm continuum, and a
\CEO\ (2--1) velocity map with the possible signature of a rotating
disk.  For the analysis of Per-8, fits to the observed spectra were
made only for those profiles that had three independent velocity
channels with a signal-to-noise ratio greater than 5.  In Figure
\ref{SPECTRA}, example spectra of \CEO\ (2-1) and \CSO\ (2-1) at the
peak of the line emission are shown.  In Figure \ref{IRACMAP}, we show
that the 1.3\,mm continuum emission presented here is spatially
coincident with the IRAC source noted by \citet{Sadavoy10b} and the
peak of the SCUBA 850\,\micron\ emission.

Per-8 shows that some cores cannot be accurately classified solely
from infrared continuum maps and assumed protostellar colors.  For
example, strong PAH emission or shocks from outflows may cause excess
emission in some of the IRAC bands and skew the infrared colors
\citep[e.g.,][]{Gutermuth08}.  Assumed protostellar colors are
determined based on a best-effort basis, and only select objects {\it
  most likely} to be protostars.  Furthermore, the photometry of Per-8
is complicated by it being a resolved source and by its location on
the boundary between two c2d mosaic tiles.  Occasionall misclassified
cores like Per-8 should be expected and caution is necessary when
using only infrared continuum maps to determine if cores are starless
or protostellar.

\subsection{Oph-1, Oph-2, Per-2, and Per-6} \label{NONDETECTIONS}

The 1.3\,mm continuum emission from the remaining 4 cores in this
survey is not detected by our SMA and CARMA observations.  We
calculate upper limits to the masses of point sources that could have
been detected in these observations from the respective rms values
reported in Table \ref{OBSPROPTABLE} and using Equation \ref{DUSTMASS}
scaled to 1.3\,mm with an assumed emissivity spectral index of $\beta
= 2$.  The 3\,$\sigma$ upper limits on the mass of a point source
embedded in Oph-1, Oph-2, Per-2, and Per-6 are 0.002\,\solmass,
0.002\,\solmass, 0.01\,\solmass, and 0.02\,\solmass\ respectively.
Had we assumed a lower dust temperature of 5.5\,K, as reported by
\citet{Crapsi07} at the center of L1544, our reported mass upper
limits would be increased by a factor of 4-5.  In comparison, the
median 3\,$\sigma$ upper limit of the mass of compact structures in
starless cores in Perseus reported by \citet{Schnee10} is
0.2\,\solmass, so this survey goes a factor of 10-100 deeper than
\citet{Schnee10}.  The improved mass sensitivity in this paper is
primarily a result of the wavelengths of observations \citep[1.3\,mm
  here vs.~3\,mm by][]{Schnee10} and the $\lambda^{-4}$ dependence of
the dust emission.  We conclude that Oph-1, Oph-2, Per-2, and Per-6
are true starless cores, based on the non-detection of compact
continuum down to a few Jupiter masses.  In addition to being
undetected in the 1.3\,mm continuum maps, Oph-1, Oph-2, Per-2, and
Per-6 show no signs of outflows in the CO and \TCO\ maps.  The
non-detection of outflows from these cores further suggests that they
are indeed starless.  Furthermore, these four cores are not detected
in the \NTDP, \CEO, or \CSO\ line maps made with the SMA and CARMA.

\subsubsection{Density Profiles}

\begin{figure}
\epsscale{1.0} 
\plotone{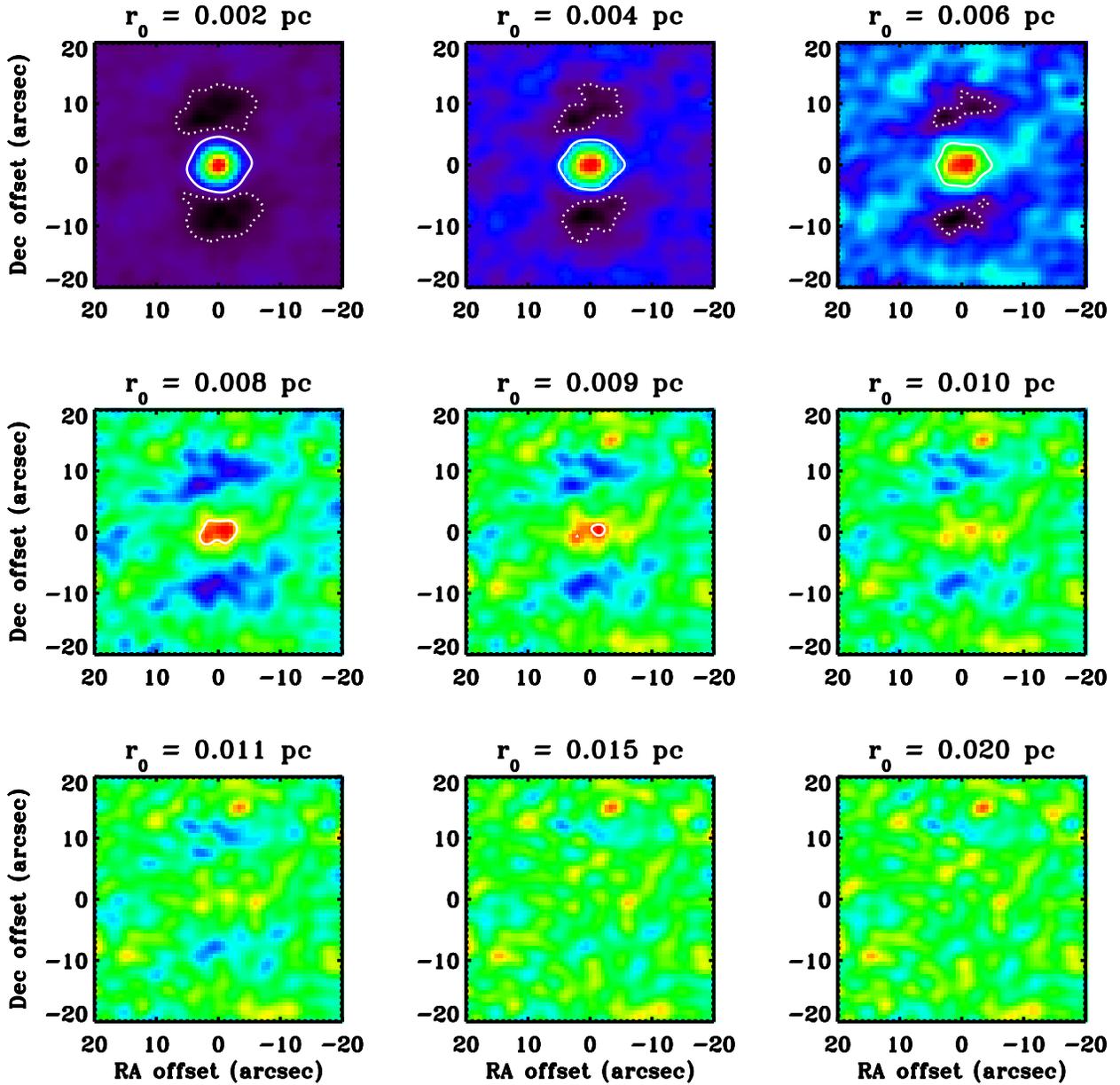}
\caption{Simulated SMA compact array maps of starless cores of
  constant temperature ($T_d$ = 10\,K) in the Perseus molecular cloud.
  The observations are described in Section \ref{NONDETECTIONS}.  The
  radius inside of which the density profile is flat is labeled for
  each panel.  The thermal noise in each map is 0.5\,mJy\,beam$^{-1}$.
  The solid contours show emission at 2.5\,mJy\,beam$^{-1}$ and the
  dashed contours show emission at -2.5\,mJy\,beam$^{-1}$.  There is
  no $\pm$5\,$\sigma$ emission for $r_0 \ge 0.01$\,pc.  Regions of
  negative flux are artefacts originating from incomplete sampling of
  the Fourier transform of the sky brightness distribution.
\label{SIMFIG}}
\end{figure}

To analyze further the meaning of non-detections of continuum emission
from starless cores, we simulate observations of the 1.3\,mm dust
emission from idealized cores.  We assumed a typical density
distribution \citep{Tafalla04} of
\begin{equation}
n = \frac{n_0}{1 + \left(\frac{r}{r_0}\right)^{2.5}}
\end{equation}
where $n_0$ is normalized such that the total core mass within a
radius of 0.05\,pc is 5\,\solmass.  This density profile has a flat
interior (at radii less than $r_0$) and a steep exterior (at radii
greater than $r_0$).  The radius $r_0$ varies from 0.002\,pc to
0.02\,pc in our simulations, as shown in Figure \ref{SIMFIG}.  We used
an outer radius of 0.05\,pc, as is typical of starless cores.  We
assumed a dust emissivity at 1.3\,mm of 0.009\,cm$^2$\,g$^{-1}$, as
given in Table 1, column 5 of \citet{Ossenkopf94} for dust grains with
thin ice mantles at a density of 10$^6$\,cm$^{-3}$.  We assumed a
distance to the cores of 250\,pc, appropriate for cores in the Perseus
molecular cloud \citep{Hirota08, Lombardi10}.  In one set of
simulations, we assumed an isothermal dust temperature of 10\,K,
appropriate for starless cores in Perseus \citep{Schnee09}.  In a
second set of simulations, we assumed the dust temperature profile
derived for the starless core L1544 by \citet{Crapsi07}.
\begin{equation}
\label{TREQ}
T(r) = T_{out} - \frac{T_{out} - T_{in}}{1 + (r/r_{0,T})^{1.5}}
\end{equation}
where $T_{out}$ is the temperature in the outer layer of the core and
had a value of 12\,K, $T_{in}$ is the temperature at the center of the
core and had a value of 5.5\,K, and $r_{0,T}$ had a value of
0.012\,pc.

The 2-dimensional flux distribution was predicted for each core, and
this was used as input for the simulated observations with SMA.  We
used the Common Astronomy Software Applications
(CASA)\footnote{http://casa.nrao.edu} software package to simulate a
5-hour track on each core, centered on transit, with thermal noise
equal to 0.5\,mJy\,beam$^{-1}$, which is similar to the observations
described in Section \ref{SMA}.

\begin{figure}
\epsscale{1.0} 
\plotone{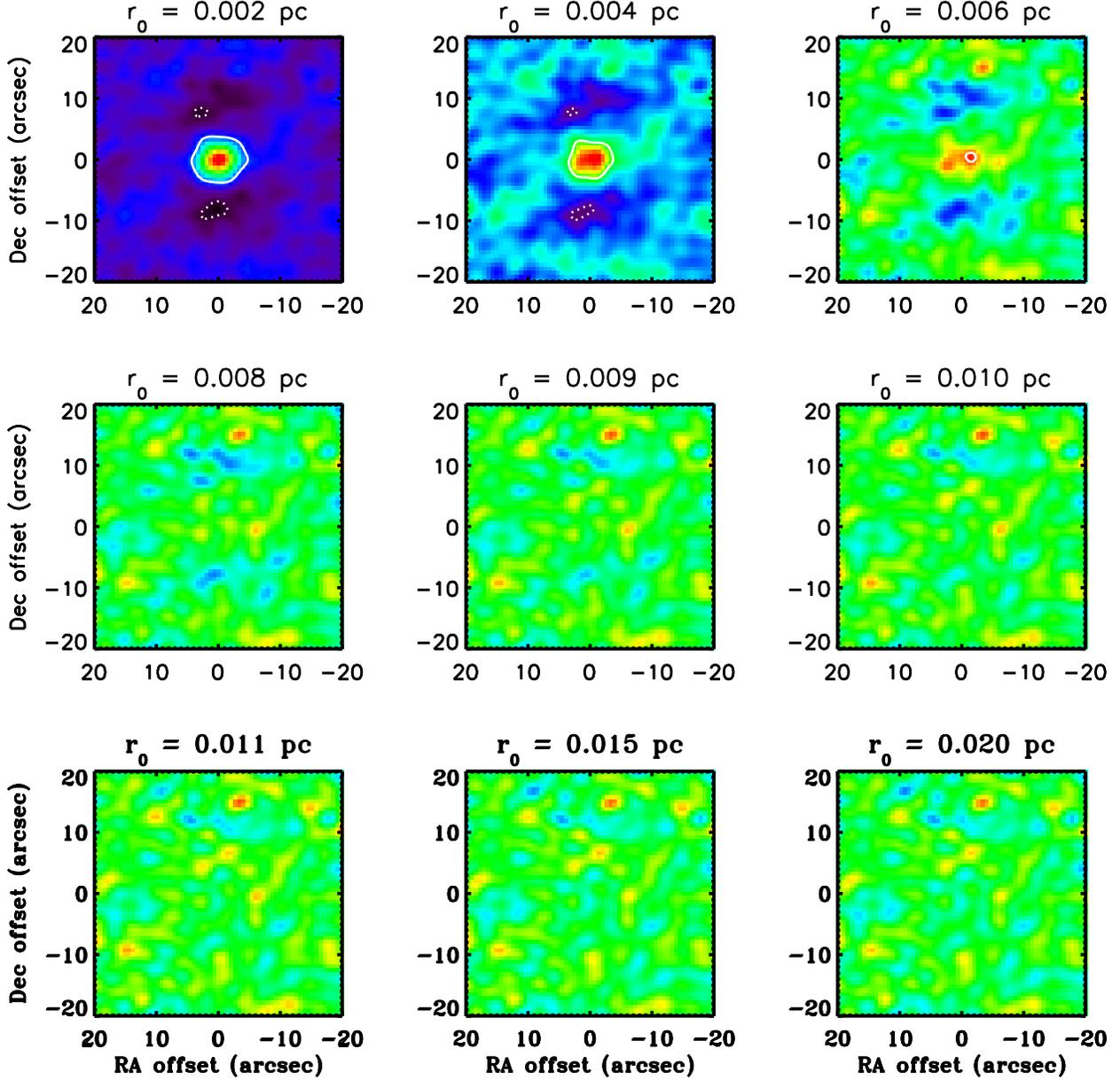}
\caption{Same as Figure \ref{SIMFIG}, but for cores with a temperature
  profile given by Equation \ref{TREQ}.  In comparison with Figure
  \ref{SIMFIG}, the only cores that are detected in these simulations
  with at least 5\,$\sigma$ confidence are those with $r_0 \le
  0.006$\,pc.  There is no $\pm$5\,$\sigma$ emission for $r_0 \ge
  0.008$\,pc.
\label{SIMFIGTR}}
\end{figure}

As shown in Figure \ref{SIMFIG}, simulated isothermal cores with more
compact density distributions ($r_0 \le 0.009$\,pc) would have been
detected with 5\,$\sigma$ confidence, while cores with larger central
flat regions ($r_0 > 0.009$\,pc) are resolved out by the simulated SMA
observations.  To be consistent with the observations presented here,
cores with a more realistic temperature profile and small central flat
density profile ($r_0 \le 0.006$\,pc) are ruled out, as shown in
Figure \ref{SIMFIGTR}.

Since we do not detect any of the starless cores in our sample, we
estimate that their density profiles are flat on scales out to at
least 0.01\,pc, assuming isothermality.  Allowing for an expected
temperature drop toward the core center, we find that these cores must
be flat out to scales larger than 0.006\,pc.  The density profiles
that we find are in excellent agreement with previous single-dish
observations.  For example, \citet{Ward-Thompson99} mapped with the
IRAM 30\,m telescope the 1.3\,mm continuum emission from a sample of
nearby starless cores and found that their flux distributions were
flat out to 0.01-0.02\,pc.  The smoothness of the continuum emission
from starless cores found in this study is also in agreement with the
results of \citet{Schnee10} and \citet{Offner12}.

\subsubsection{Collapse Rate}

Based on a simple Jeans analysis, these starless cores may be unstable
and therefore in the process of collapsing.  The non-detection of
continuum emission from the starless cores in our sample, however, is
inconsistent with collapse models that predict strongly peaked density
distributions, such as those seen in expansion wave models
\citep{Silk88}.  The early stages of slow contraction, such as in
models of ambipolar diffusion \citep[e.g.,][]{Safier97} and models of
the pressure-free collapse of Bonnor-Ebert spheres
\citep[e.g.,][]{Myers05}, have flat density distributions at small
radii and are consistent with our observations.  Similarly, the
initial stages of uniform collapse \citep{Silk88} are also consistent
with our observations.  The infall rate, in a freefall approximation,
for a 5\,\solmass\ object to collapse from an initial radius of
0.1\,pc to 0.05\,pc (roughly the radius of the cores in this sample)
is approximately 0.2\,\kms.  Collapse at this rate can be easily
detected in starless cores, as the typical line width is comparable to
this infall speed \citep{Foster09}. We suggest that super-Jeans
starless cores are good targets for future observations to look for
infall.

\subsubsection{Stability Against Collapse}

The reported values of $M/M_J$ for Oph-1, Oph-2, Per-2 and Per-6 are
2.2, 2.3, 4.8 and 4.9 \citep{Sadavoy10b} respectively, for a Jeans
mass derived from thermal support only.  Although the starless cores
in this paper are super-Jeans when considering only thermal support,
we can estimate their stability when also accounting for their
non-thermal motions.  For this analysis, we looked at archival data of
the published velocity dispersions for these cores.
\citet{Rosolowsky08} have published the NH$_3$ (1,1) velocity
dispersions of Per-2 and Per-6, i.e., 0.31\,\kms\ and
0.46\,\kms\ respectively.  The N$_2$H$^+$ (1-0) velocity dispersions
of Per-2 and Per-6, 0.33\,\kms\ and 0.32\,\kms, are similar to the
ammonia line dispersions \citep{Johnstone10}.  The typical velocity
dispersion measured in NH$_3$ spectra of starless cores in Perseus is
$<$0.2\,\kms\ \citep{Foster09}, so Per-2 and Per-6 have especially
large non-thermal motions.  The velocity dispersion due to thermal
motions NH$_3$ at 10\,K is $\sim$0.07\,\kms, or about 4-6$\times$
smaller than the observed dispersions measured from ammonia spectra of
Per-2 and Per-6.  \citet{Friesen09} and \citet{Roueff05} have
published the NH$_3$ and ND$_3$ velocity dispersions for regions near
Oph-1 and Oph-2, which have values of 0.16\,\kms\ and 0.18\,\kms,
about 2 times larger than the thermal velocity dispersion.

To determine the stability of the starless cores in our sample,
including the non-thermal motions, we calculate a ``typical'' virial
mass for a uniform density core, following \citet{Bertoldi92}.

\begin{equation}
M_{\rm vir} = \frac{5R\sigma^2}{G}
\end{equation} 

Taking values typical of the starless cores in this sample ($\sigma =
$0.4\,\kms\ and $R = $0.05\,pc), we find that a typical virial mass is
$\sim$9\,\solmass\ (or roughly 4.5\,\solmass\ for a density
distribution of a critically stable Bonnor-Ebert sphere).  Given that
the actual masses of the cores are approximately equal to the
Bonnor-Ebert virial mass, we can not be completely confident that the
cores are bound or gravitationally unstable.  The uncertainties in the
true core masses and non-thermal support, however, are significant.
Indeed, the large line widths seen in these cores could be interpreted
as coming from a structured velocity field (from infall itself, for
example), and may not be an indication of turbulent support at all.
We note also that the manner in which the observed non-thermal motions
might supply pressure support without significant dissipation is still
unclear.  High angular resolution observations of the velocity fields
around these cores can distinguish between these possibilities.

\section{Summary} \label{SUMMARY}

In this paper, we presented new SMA and CARMA observations of five
super-Jeans cores that have been previously classified as being
candidate ``starless'' or ``undetermined'' by \citet{Sadavoy10b}.  We
find, in agreement with previous observations, that the core Per-8 is
actually protostellar and is the origin of a collimated bipolar
outflow.  The cores Oph-1, Oph-2, Per-2, and Per-6 are truly starless,
however, with non-detections in the 1.3\,mm continuum corresponding to
3\,$\sigma$ upper limits to the mass of any embedded point source of
$\le$0.02\,\solmass.  Our simulations suggest that these four cores
have flat density profiles out to radii of at least
$\sim$0.006--0.01\,pc, in agreement with previous single-dish and
interferometric observations of starless cores.  Although Oph-1, Oph2,
Per-2, and Per-6 are super-Jeans when considering only their thermal
support, when non-thermal support is considered their stability
against collapse is much more uncertain.

\acknowledgments

We thank Robert Gutermuth for helpful discussion on the nature of
Per-8.  We thank our anonymous referee for suggestions that improved
this paper.  The National Radio Astronomy Observatory is a facility of
the National Science Foundation operated under cooperative agreement
by Associated Universities, Inc.  JDF acknowledges support by the
National Research Council of Canada and the Natural Sciences and
Engineering Council of Canada (via a Discovery Grant).  DJ is
supported by a Natural Sciences and Engineering Research Council of
Canada (NSERC) Discovery Grant.  The Submillimeter Array is a joint
project between the Smithsonian Astrophysical Observatory and the
Academia Sinica Institute of Astronomy and Astrophysics and is funded
by the Smithsonian Institution and the Academia Sinica.  Support for
CARMA construction was derived from the Gordon and Betty Moore
Foundation, the Kenneth T. and Eileen L. Norris Foundation, the James
S. McDonnell Foundation, the Associates of the California Institute of
Technology, the University of Chicago, the states of California,
Illinois, and Maryland, and the National Science Foundation. Ongoing
CARMA development and operations are supported by the National Science
Foundation under a cooperative agreement, and by the CARMA partner
universities.

{\it Facilities}: SMA, CARMA

{}

\begin{deluxetable}{cccccccc} 
\tablewidth{0pt}
\tabletypesize{\scriptsize}
\tablecaption{Summary of Dense Core Properties\label{PROPERTIESTABLE}}
\tablehead{
 \colhead{Name\tablenotemark{1}}  & 
 \colhead{Nearby Sources \tablenotemark{1}}  &
 \colhead{RA}                     &
 \colhead{Dec}                    & 
 \colhead{Molecular Cloud}        &
 \colhead{Distance}               & 
 \colhead{Mass\tablenotemark{1}}  &
 \colhead{R$_{\rm eff}$\tablenotemark{1}} \\ 
 \colhead{}                       &
 \colhead{}                       &
 \colhead{J2000}                  &
 \colhead{J2000}                  &
 \colhead{}                       &
 \colhead{pc}                     &
 \colhead{\solmass}               &
 \colhead{pc}}
\startdata
  Oph-1 & ...              & 16:26:59.20 & -24:34:20.0 & Ophiuchus & 125 & 5.3 & 0.060\\
  Oph-2 & IRAS 16293-2422E & 16:32:29.00 & -24:29:06.0 & Ophiuchus & 125 & 3.2 & 0.038\\
  Per-2 & ...              & 03:28:59.50 & +31:21:31.0 & Perseus   & 250 & 7.6 & 0.053\\
  Per-6 & ...              & 03:29:08.70 & +31:15:13.0 & Perseus   & 250 & 8.1 & 0.055\\
  Per-8 & IRAS 03292-3039  & 03:32:17.60 & +30:49:47.0 & Perseus   & 250 & 7.2 & 0.055\\
\enddata
\tablenotetext{1}{From \citet{Sadavoy10b}}
\end{deluxetable}

\begin{deluxetable}{ccccccc} 
\tablewidth{0pt}
\tabletypesize{\scriptsize}
\tablecaption{SMA and CARMA Map Properties\label{OBSPROPTABLE}}
\tablehead{
 \colhead{Name}                             & 
 \colhead{Beam size\tablenotemark{1}}       & 
 \colhead{Continuum rms\tablenotemark{1,2}} &
 \colhead{\CEO\ rms\tablenotemark{1,3}}     &
 \colhead{Beam size\tablenotemark{4}}       & 
 \colhead{Continuum rms\tablenotemark{4,2}} &
 \colhead{\CSO\ rms\tablenotemark{4,5}}     \\
 \colhead{}                                 &
 \colhead{\arcsec}                          &
 \colhead{mJy beam$^{-1}$}                   &
 \colhead{mJy beam$^{-1}$}                   &
 \colhead{\arcsec}                          &
 \colhead{mJy beam$^{-1}$}                   &
 \colhead{mJy beam$^{-1}$}}
\startdata
  Oph-1 & 3.3 $\times$ 2.4 & 0.54 & 150 &                  &     &     \\
  Oph-2 & 3.3 $\times$ 2.4 & 0.63 & 150 &                  &     &     \\
  Per-2 & 3.7 $\times$ 2.4 & 0.61 & 120 & 2.2 $\times$ 1.7 & 2.4 & 73  \\
  Per-6 & 3.7 $\times$ 2.4 & 0.75 &  72 &                  &     &     \\
  Per-8 & 3.7 $\times$ 2.4 & 2.5  &  69 & 2.2 $\times$ 1.8 & 8.8 & 79  \\
\enddata
\tablenotetext{1}{SMA observations}
\tablenotetext{2}{Over the 7\,GHz band width}
\tablenotetext{3}{Per 0.28\,\kms\ channel}
\tablenotetext{4}{CARMA observations}
\tablenotetext{5}{Per 1\,\kms\ channel}
\end{deluxetable}

\begin{deluxetable}{llccc} 
\tablewidth{0pt}
\tabletypesize{\scriptsize}
\tablecaption{Observations \label{OBSTABLE}}
\tablehead{
 \colhead{Observatory}            &
 \colhead{Continuum/Line}         & 
 \colhead{$\nu$\tablenotemark{1}} &
 \colhead{Channel Spacing}        &
 \colhead{Band width}             \\ 
 \colhead{}                       &
 \colhead{}                       &
 \colhead{GHz}                    &
 \colhead{kHz}                    &
 \colhead{MHz}}
\startdata
  SMA & 1.3\,mm continuum & 224.538 & 812 & 7000 \\
  SMA & CO (2-1)          & 230.538 & 406 & 104 \\
  SMA & \TCO\ (2-1)       & 220.399 & 406 & 104 \\
  SMA & \CEO\ (2-1)       & 219.560 & 203 & 104 \\
  SMA & \NTDP (3-2)       & 231.322 & 812 & 104 \\
  CARMA & 1.3\,mm continuum & 227.638 & 812 & 7000 \\
  CARMA & CO (2-1)          & 230.538 & 781 & 124 \\
  CARMA &\CSO\ (2-1)       & 224.714 & 781 & 124 \\
\enddata
\tablenotetext{1}{Rest frequency}
\end{deluxetable}

\begin{thebibliography}{}
\bibitem[Adams et al.(1989)]{Adams89} Adams, F.~C., Ruden, S.~P., \&
  Shu, F.~H.\ 1989, \apj, 347, 959
\bibitem[Bertoldi \& McKee(1992)]{Bertoldi92} Bertoldi, F., \& McKee,
  C.~F.\ 1992, \apj, 395, 140
\bibitem[Bonnell \& Bate(1994)]{Bonnell94} Bonnell, I.~A., \& Bate,
  M.~R.\ 1994, \mnras, 269, L45
\bibitem[Chen \& Arce(2010)]{Chen10} Chen, X., \& Arce, H.~G.\ 2010,
  \apjl, 720, L169
\bibitem[Chen et al.(2012)]{Chen12} Chen, X., Arce, H.~G., Dunham,
  M.~M., et al.\ 2012, arXiv:1203.5252
\bibitem[Corder et al.(2010)]{Corder10} Corder, S.~A., Wright,
  M.~C.~H., \& Carpenter, J.~M.\ 2010, \procspie, 7733,
\bibitem[Crapsi et al.(2007)]{Crapsi07} Crapsi, A., Caselli, P.,
  Walmsley, M.~C., \& Tafalla, M.\ 2007, \aap, 470, 221
\bibitem[Di Francesco et al.(2008)]{DiFrancesco08} Di Francesco, J.,
  Johnstone, D., Kirk, H., MacKenzie, T., \& Ledwosinska, E.\ 2008,
  \apjs, 175, 277
\bibitem[Dunham et al.(2011)]{Dunham11} Dunham, M.~M., Chen, X., Arce,
  H.~G., et al.\ 2011, \apj, 742, 1
\bibitem[Enoch et al.(2010)]{Enoch10} Enoch, M.~L., Lee, J.-E.,
  Harvey, P., Dunham, M.~M., \& Schnee, S.\ 2010, \apjl, 722, L33
\bibitem[Evans et al.(2003)]{Evans03} Evans, N.~J., II, Allen, L.~E.,
  Blake, G.~A., et al.\ 2003, \pasp, 115, 965
\bibitem[Evans et al.(2009)]{Evans09} Evans, N.~J., II, Dunham, M.~M.,
  J{\o}rgensen, J.~K., et al.\ 2009, \apjs, 181, 321
\bibitem[Fisher(2004)]{Fisher04} Fisher, R.~T.\ 2004, \apj, 600, 769
\bibitem[Foster et al.(2009)]{Foster09} Foster, J.~B., Rosolowsky,
  E.~W., Kauffmann, J., et al.\ 2009, \apj, 696, 298
\bibitem[Friesen et al.(2009)]{Friesen09} Friesen, R.~K., Di
  Francesco, J., Shirley, Y.~L., \& Myers, P.~C.\ 2009, \apj, 697,
  1457
\bibitem[Goodwin et al.(2004)]{Goodwin04} Goodwin, S.~P., Whitworth,
  A.~P., \& Ward-Thompson, D.\ 2004, \aap, 414, 633
\bibitem[Goodwin et al.(2007)]{Goodwin07} Goodwin, S.~P., Kroupa, P.,
  Goodman, A., \& Burkert, A.\ 2007, Protostars and Planets V, 133
\bibitem[Gutermuth et al.(2008)]{Gutermuth08} Gutermuth, R.~A.,
  Bourke, T.~L., Allen, L.~E., et al.\ 2008, \apjl, 673, L151
\bibitem[Hatchell \& Dunham(2009)]{Hatchell09} Hatchell, J., \&
  Dunham, M.~M.\ 2009, \aap, 502, 139
\bibitem[Hirota et al.(2008)]{Hirota08} Hirota, T., Bushimata, T.,
  Choi, Y.~K., et al.\ 2008, \pasj, 60, 37
\bibitem[Ho et al.(2004)]{Ho04} Ho, P.~T.~P., Moran, J.~M., \& Lo,
  K.~Y.\ 2004, \apjl, 616, L1
\bibitem[Johnstone et al.(2010)]{Johnstone10} Johnstone, D.,
  Rosolowsky, E., Tafalla, M., \& Kirk, H.\ 2010, \apj, 711, 655
\bibitem[Kirk et al.(2009)]{Kirk09} Kirk, J.~M., Crutcher, R.~M., \&
  Ward-Thompson, D.\ 2009, \apj, 701, 1044
\bibitem[Ladd et al.(1998)]{Ladd98} Ladd, E.~F., Fuller, G.~A., \&
  Deane, J.~R.\ 1998, \apj, 495, 871
\bibitem[Lombardi et al.(2010)]{Lombardi10} Lombardi, M., Lada, C.~J.,
  \& Alves, J.\ 2010, \aap, 512, A67
\bibitem[Myers(2005)]{Myers05} Myers, P.~C.\ 2005, \apj, 623, 280
\bibitem[Offner et al.(2012)]{Offner12} Offner, S.~S.~R., 
Capodilupo, J., Schnee, S., \& Goodman, A.~A.\ 2012, \mnras, 420, L53 
\bibitem[Olmi et al.(2005)]{Olmi05} Olmi, L., Testi, L., \& Sargent,
  A.~I.\ 2005, \aap, 431, 253
\bibitem[Ossenkopf \& Henning(1994)]{Ossenkopf94} Ossenkopf, V., \&
  Henning, T.\ 1994, \aap, 291, 943
\bibitem[Pineda et al.(2011)]{Pineda11} Pineda, J.~E., Arce, H.~G.,
  Schnee, S., et al.\ 2011, \apj, 743, 201
\bibitem[Rosolowsky et al.(2008)]{Rosolowsky08} Rosolowsky, E.~W.,
  Pineda, J.~E., Foster, J.~B., et al.\ 2008, \apjs, 175, 509
\bibitem[Roueff et al.(2005)]{Roueff05} Roueff, E., Lis, D.~C., van
  der Tak, F.~F.~S., Gerin, M., \& Goldsmith, P.~F.\ 2005, \aap, 438,
  585
\bibitem[Sadavoy et al.(2010b)]{Sadavoy10b} Sadavoy, S.~I., Di
  Francesco, J., \& Johnstone, D.\ 2010b, \apjl, 718, L32
\bibitem[Sadavoy et al.(2010a)]{Sadavoy10a} Sadavoy, S.~I., Di
  Francesco, J., Bontemps, S., et al.\ 2010a, \apj, 710, 1247
\bibitem[Safier et al.(1997)]{Safier97} Safier, P.~N., McKee, C.~F.,
  \& Stahler, S.~W.\ 1997, \apj, 485, 660
\bibitem[Sault et al.(1995)]{Sault95} Sault, R.~J., Teuben, P.~J., \&
  Wright, M.~C.~H.\ 1995, Astronomical Data Analysis Software and
  Systems IV, 77, 433
\bibitem[Schnee et al.(2009)]{Schnee09} Schnee, S., Rosolowsky, E.,
  Foster, J., Enoch, M., \& Sargent, A.\ 2009, \apj, 691, 1754
\bibitem[Schnee et al.(2010)]{Schnee10} Schnee, S., Enoch, M.,
  Johnstone, D., et al.\ 2010, \apj, 718, 306
\bibitem[Schnee et al.(2012)]{Schnee12} Schnee, S., Di Francesco, J.,
  Enoch, M., et al.\ 2012, \apj, 745, 18
\bibitem[Silk \& Suto(1988)]{Silk88} Silk, J., \& Suto, Y.\ 1988,
  \apj, 335, 295
\bibitem[Tafalla et al.(2004)]{Tafalla04} Tafalla, M., Myers, P.~C.,
  Caselli, P., \& Walmsley, C.~M.\ 2004, \aap, 416, 191
\bibitem[Walawender et al.(2009)]{Walawender09} Walawender, J., 
Reipurth, B., \& Bally, J.\ 2009, \aj, 137, 3254 
\bibitem[Ward-Thompson et al.(1999)]{Ward-Thompson99} Ward-Thompson,
  D., Motte, F., \& Andre, P.\ 1999, \mnras, 305, 143

\end{thebibliography}
\end{document}